# Intuitive Control of Scraping and Rubbing Through Audio-tactile Synthesis

Mitsuko Aramaki, Corentin Bernard, Richard Kronland-Martinet, Samuel Poirot, Sølvi Ystad

Aix Marseille Univ, CNRS, PRISM (Perception, Representations, Image, Sound, Music),
31 Chemin J. Aiguier, 13402 Marseille Cedex 20, France

`{name}@prism.cnrs.fr`

**Abstract.** Intuitive control of synthesis processes is an ongoing challenge within the domain of auditory perception and cognition. Previous works on sound modelling combined with psychophysical tests have enabled our team to develop a synthesizer that provides intuitive control of actions and objects based on semantic descriptions for sound sources. In this demo we present an augmented version of the synthesizer in which we added tactile stimulations to increase the sensation of true continuous friction interactions (rubbing and scratching) with the simulated objects. This is of interest for several reasons. Firstly, it enables to evaluate the realism of our sound model in presence of stimulations from other modalities. Secondly it enables to compare tactile and auditory signal structures linked to the same evocation, and thirdly it provides a tool to investigate multimodal perception and how stimulations from different modalities should be combined to provide realistic user interfaces.

## 1    Introduction

Previous results in the field of multimodal perception have provided examples of strong perceptual influences between modalities. One well-known example is the McGurk effect in which visual stimuli influence speech perception [8]. More recent studies revealed that sounds can modify the perception of a visual trajectory and even the gestural reproduction of the visual shape [12]. In the case of touch perception, several studies have revealed a strong influence of auditory stimuli on perceived textures [2, 5, 6,7, 10, 11].

In the present study we explore such multimodal interactions in the light of our previous works on intuitive sound control that describes the sound as the result of an action on an object. This approach presumes the existence of sound invariants responsible for the evocation of sound events [4], and has led to a synthesizer that makes it possible to control sounds from semantic labels that describe the action (rubbing, scratching rolling) and the object (material, shape, size, …). Continuous control between the different evocations makes it possible for the users to freely navigate between different actions hereby creating both realistic and virtual sound events [1,3,9].

As a first approach to multimodal synthesis, we focus on evocations of continuous friction interactions, in particular rubbing and scraping, to investigate whether tactile invariants of these actions exist and whether they resemble the corresponding auditory invariants. In the following section we describe the synthesis process of the tactile stimulation, the experimental setup and some preliminary results of ongoing perceptual tests.

## 2     Synthesis of Auditory and Tactile Stimulations Evoking Continuous Friction Interactions

As a first approach to investigate perceptual invariants for tactile structures, we focus on the evocation of two different continuous interactions namely scraping and rubbing. In the auditory domain it has been shown that these actions can be simulated by successive impacts (see Fig. 1) with different temporal intensities [13]. The impact distribution is smoother for rubbing than for scratching since scratching is considered as an action in which the interaction with each surface irregularity is encountered one after another and more intensely than in the case of rubbing. This model was perceptually validated by Conan et al [3] and confirmed that impact distributions are associated to the auditory invariant allowing for the distinction between scratching and rubbing.

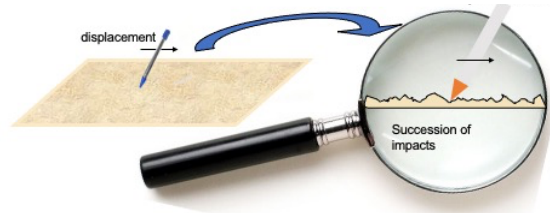

**Fig. 1.** Phenomenological model of continuous interactions

Would this also be the case in the tactile domain? To answer this question, we designed a synthesis model based on the same features as in the auditory modality, i.e. mean and standard deviations of the amplitudes and the temporal distance between successive peaks, to investigate evocations of rubbing and scratching using an actuator attached to a pen. Then we conducted a perceptual test in which subjects were asked to explore the surface of a graphic tablet and to determine (on a continuous cursor) whether the sensation evoked scraping or rubbing. The experimental protocol is described in the next section.

## 3 Perceptual Evaluations

Sixteen subjects evaluated 96 evoked continuous interactions induced by auditory and tactile stimulations. They wore anti-noise headphones when evaluating the tactile stimuli. During the tactile evaluations, they were asked to hold a pen equipped with the actuator and to explore a surface of a graphic tablet. After the exploration they evaluated the evocation on a continuous one-dimensional scale between the (French) words "gratter" (scratch) and "frotter" (rub). While the preliminary results confirmed previous findings related to the impact distribution as the most influent parameter on the evocation of rubbing and scratching in the auditory domain [3], this parameter did not turn out to have a significant influence in the tactile domain. On the other hand, amplitude variations tended to be more important in the tactile domain and had a significant influence on the perceived action. Scratching evocations were associated with strong amplitudes while the weakest amplitudes were associated with rubbing.

## 4 Audio-tactile Synthesizer

The current study suggests that perceptual invariants differ in the case of auditory and tactile perception. In the case of simulations of continuous friction interactions, temporal variations are essential in the auditory domain while amplitude variations seem to play a greater role in the tactile domain. The proposed demo consists of a multimodal synthesizer calibrated according to the previous perceptual results that enables participants to explore auditory and tactile signal invariants and to combine the evocations with auditory evocations of objects (see Fig. 2). The user is invited to wear headphones (for auditory stimulations) and to hold a pen equipped with the actuator (for tactile stimulations) coupled with a tablet. A computer displays a graphical interface on which the user can choose the type of interactions (rubbing or scratching) in a continuous way.

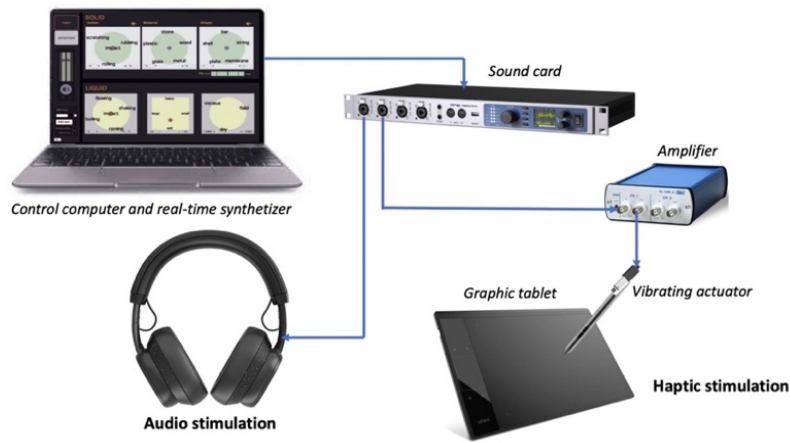

**Fig. 2.** Set up of the synthesizer device

**Acknowledgements** This work was partly finance by the French National Research Agency (ANR) in the case of the France Relance program (C. Bernard) and the COMMUTE ANR-22-CE33-0009 project and the by Institute of Language, Communication and the Brain (ILCB)/Center of Excellence on Brain and Language (BLRI) Grant Nos. ANR-16-CONV-0002 (ILCB) and ANR-11-LABX-0036 (BLRI), the Excellence Initiative of Aix- Marseille University (AMIDEX). We would like to thank Raphaël Vancheri for his precious contribution to the perceptive evaluations.